\documentclass[aps,pra,twocolumn,groupedaddress,floatfix,showpacs]{revtex4-1}
\usepackage{amssymb,amsmath}
\usepackage{graphicx}
\usepackage{subfigure}
\usepackage[english]{babel}
\usepackage{float}
\usepackage{color}

\newcommand{\be}{\begin{equation}}
\newcommand{\ee}{\end{equation}}

\begin{document}

\title{Compact modes in quasi one dimensional coupled magnetic oscillators}

\author{Dany L\'{o}pez-Gonz\'{a}lez and Mario I. Molina}

\address{Department of Physics, MSI-Nucleus on Advanced Optics, Faculty of sciences, University of Chile, Santiago, Chile}

\begin{abstract}
In this work we study analytically and numerically the spectrum and localization properties of three quasi-one-dimensional (ribbons) split-ring resonator arrays which possess magnetic flatbands, namely, the stub, Lieb and kagome lattices, and how their spectra is affected by the presence of perturbations that break the delicate geometrical interference needed for a magnetic flatband to exist. We find that the Stub and Lieb ribbons are stable against the three types of perturbations considered here, while the kagome ribbon is, in general, unstable. When losses are incorporated, all flatbands remain dispersionless but become complex, with the kagome ribbon exhibiting the highest loss rate.
\end{abstract}

\maketitle

\section{INTRODUCTION}

Magnetic Metamaterials (MMs) constitute a class of  novel artificial materials characterized for having a negative magnetic permeability, over a frequency range. A usual realization of such system consist of an array of metallic split-ring resonators (SSRs) that are coupled inductively. One of the most attractive features of SRRs is the possibility of a negative magnetic response over a given frequency region, which might overlap the frequency region where the dielectric constant of the material is negative. This gives rise to a negative index of refraction inside that frequency interval and thus, makes SRRs attractive for use as a constituent in negative refraction index materials\cite{neg_index}. Their magnetic response can be tailored to certain extent, although there are heavy Ohmmic and radiative loses. A possible solution that is to endow the SRRs with some sort of external gain, such as tunnel (Esaki) diodes\cite{esaki} to compensate for such loses. 

A periodic array of SRRs has, in principle, an energy spectrum composed of a number of bands. 
The breaking of the translational invariance, by means of impurities or disorden, gives rise to localization. Recently, attention was called to another way to achieve localization in a periodic system: Flatbands. Simply stated, a flatband lattice is a periodic system characterized by having one or more flatbands in its spectrum. Since the group velocity of a state belonging to one of these bands is zero, any flatband eigenstate or a superposition of them will exhibit no mobility. 
This allows for the formation of compacton-like structures, which are completely localized in space, exhibiting no dynamical evolution thus, constituting a new form of localized state in the continuum\cite{18}. Some systems where flatbands have been studied and observed include optical\cite{2,3} and photonic lattices\cite{4,5,6}, graphene\cite{7,8}, superconductors\cite{9,10,11,12}, fractional quantum Hall systems\cite{13,14,15}, and exciton-polariton condensates\cite{16,17}. The origin of the flatbands states can be traced back to an exact geometrical interference condition. 

One ponders what would happen if the geometrical interference condition is not completely satisfied. Would the flatbands and their accompaning phenomenology survive ? The answer to this interesting question is the main focus of this work. More specifically, we examine analytically and numerically the spectral properties of three quasi-one-dimensional (ribbons) split-ring resonator lattices: stub, Lieb and kagome lattices, which are characterized for containing flatbands in their spectrum. 
%FIG1%%%%%%%%%%%%%%%%%%%%%%%%%%%%%%%%
\begin{figure*}[t]  
\begin{center}
\includegraphics[width=14cm]{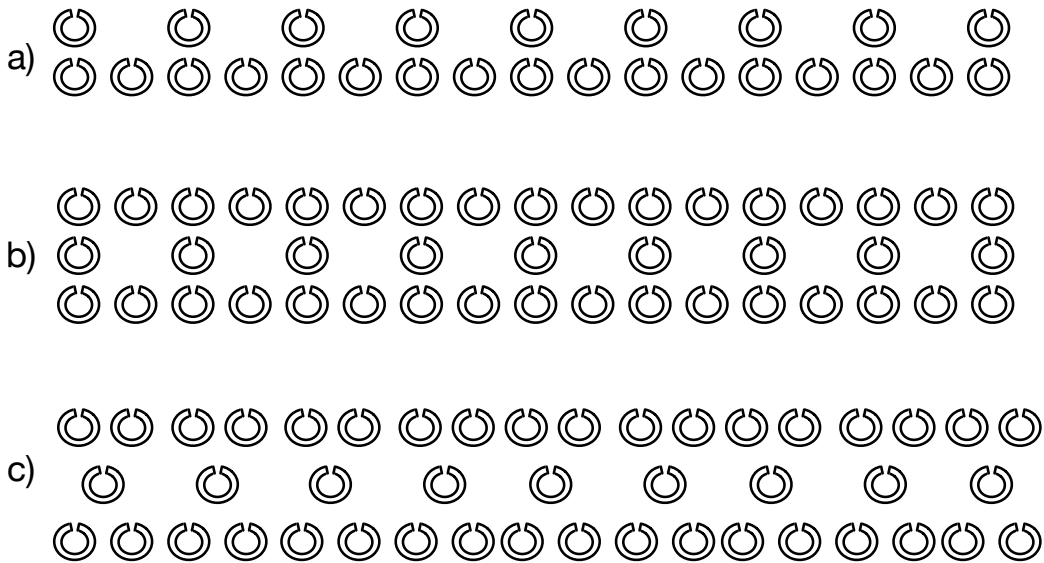}
\end{center}
\caption{Quasi-onedimensional arrays of SRRs. a) Stub ribbon b)Lieb ribbon c) Kagome ribbon.
\label{fig:Arreglos} }
\end{figure*}
%%%%%%%%%%%%%%%%%%%%%%%%%%%%%%%%%%%%%%%%%%%%%
\section{The model}
The simplest MM model consists of a quasi-onedimensional lattice (ribbon) of metallic split-ring resonators (SRRs), coupled inductively\cite{mutua}, in the absence of dissipation, driving and nonlinearity (Fig.\ref{fig:Arreglos}). Each SRR can be thought of as an approximate RCL circuit and thus, possessing a capacitance, an inductance and a resistence. They are further characterized by a resonance frequency $\displaystyle \omega_0\sim 1/\sqrt{LC}$. 

The evolution of the charge $Q$ residing on the ${\bf n}$th-ring is given by
\be
\frac{d^2}{d t^2}\left(LQ_{\bf n}+\sum_{{\bf m}\neq {\bf n}} M_{{\bf n} {\bf m}}Q_{\bf m}\right)+\frac{Q_{\bf n}}{C}=0,
\label{eq:1}
\ee
where $L$ is the self-induction of the nth-ring and $M_{{\bf n} {\bf m}}$ is the mutual inductance between rings ${\bf n}$ and ${\bf m}$. We can express this in dimensionless form by defining 
$\tau\equiv\omega_0\ t$, $q_{\bf n}\equiv Q_{\bf n}/CU_0$, $\lambda_{{\bf n} {\bf m}}\equiv M_{{\bf n} {\bf m}}/L$, with $U_0$ a characteristic voltage across the slit of the ring. Equation (\ref{eq:1}) becomes
\be
\frac{d^2}{d\tau^2}\left(q_{\bf n}(\tau)+\sum_{{\bf m}\neq {\bf n}} \lambda_{{\bf n} {\bf m}}q_{\bf m}(\tau)\right)+q_{\bf n}(\tau)=0.
\label{eq:2}
\ee
Here, $\lambda_{{\bf n} {\bf m}}$ represents the ratio between the mutual and the self inductance of the rings, and its value depends on the precise geometry of the rings as well as on the mutual distance between nearby rings. Hereafter, and for the sake of simplicity we will consider coupling to nearest-neighbors only:  $\lambda_{{\bf n},{\bf m}}=\lambda$ if ${\bf n},{\bf m}$ are nearest neighbors, zero otherwise.

The stationary modes are obtaining by posing a solution of the form $q_{{\bf n}}(\tau)=q_{{\bf n}}\exp(i \Omega \tau)$, where the $q_{{\bf n}}$ amplitudes obey
\be
-\Omega^2\left(q_{\bf n}+\lambda \sum_{n n}q_{\bf m}\right)+q_{\bf n}=0,
\label{eq:Estacionaria}
\ee
where the sum is restricted to nearest neighbors only. Since each ribbon is a an arrangement of unitary cells periodic in the horizontal direction, we set in Eq.(\ref{eq:Estacionaria})
$q_{j} = A_{j}\exp(i k n)$ where $n$ is the position of the unitary cell and $j$ labels the rings inside the unitary cell.
Using this form into Eq.(\ref{eq:Estacionaria}) leads to a $M\times M$ system of equations for the $A_{j}$ amplitudes, where $M$ is the number of rings inside a unitary cell. After imposing that the determinant of this system be zero in order to have nontrivial solutions, one arrives to a polynomial equation for $\Omega^2$ whose solutions gives us the shape of the $M$ allowed bands $\Omega^{2}(k)$.

\section{The stub ribbon}
\label{stub}
\begin{figure}[h]
\includegraphics[width=8.5cm]{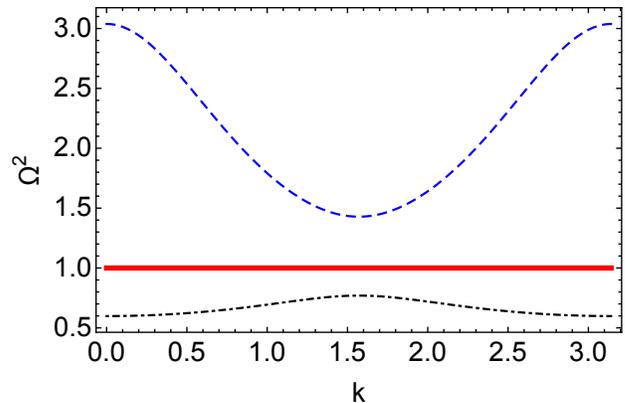}
\vspace{-0.5cm}
\caption{ The bands for the stub ribbon ($\lambda=0.3$).
\label{fig:Stubandas} }
\end{figure}

We begin by examining the stationary modes of the stub ribbon, whose geometry is shown on Fig.\ref{fig:Arreglos}a. The ribbon has a unitary cell that contains three rings, implying an spectrum with three bands. They are given by 
\begin{eqnarray}
\Omega^2 &=& 1\nonumber\\
\Omega^2&=&\frac{1}{1 \pm \sqrt{3 +2\cos(2 k)}\ \lambda}
\end{eqnarray}
The condition $\Omega^2>0$, leads to the condition
$\displaystyle |\lambda|<1/\sqrt{5}=0.447$. Under this constraint, we have all three bands real. Figure \ref{fig:Stubandas} shows the three bands.
%%%%%%%%%%%%%%%%%%%%%%%%%%%%%%%%%%
%%%%%%%%%%%%%%%%%%%%%%%%%%%%%%%%%

\begin{figure}[H]  
\begin{center}
\includegraphics[width=6cm]{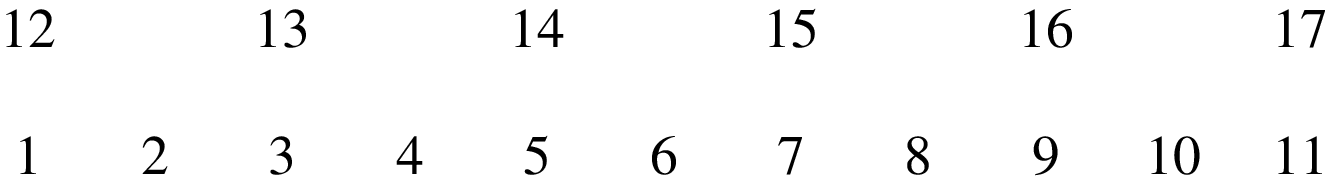}
\vspace{0.0cm}
\includegraphics[width=8cm]{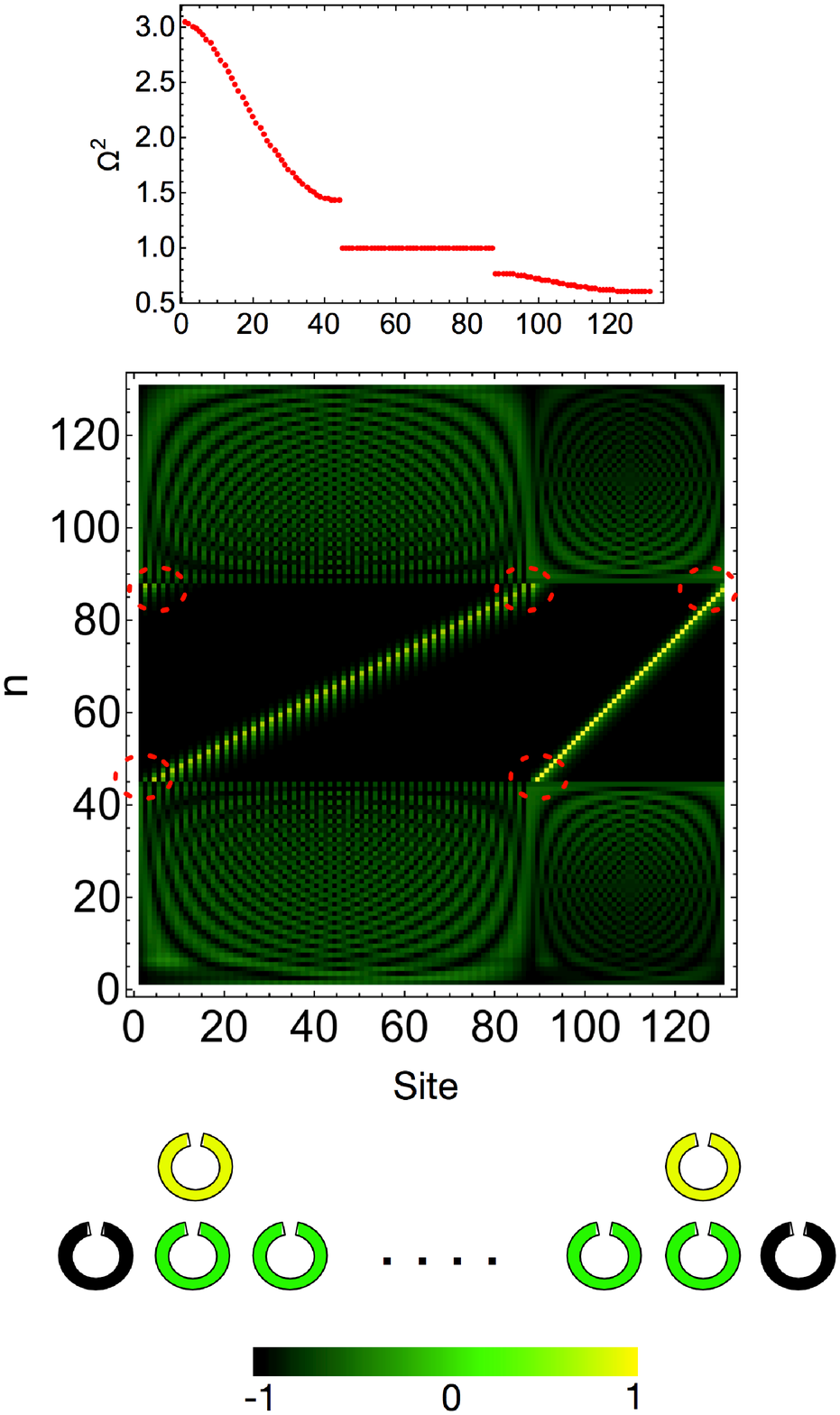}
\end{center}
\caption{Top: Eigenvalues of the Stub ribbon, arranged in order of decreasing value. The flat band is clearly appreciable at $\Omega=1$. The top part shows the numbering scheme used to label the sites. Center: Density plot of the eigenvectors of the Stub ribon, arranged in order of increasing eigenvalue. Bottom: Form of the edge modes marked in red\label{fig:modos_stub_chokley.eps} }
\end{figure}

After solving numerically the stationary equation (\ref{eq:Estacionaria}) for the stub geometry, we obtain all the modes for this ribbon. Results are shown in Fig.  \ref{fig:modos_stub_chokley.eps}. The top part shows all the eigenvalues ordered in descending order. The flat band is clearly visible. The center part shows all the modes in space, where we have placed each eigenvector in correspondence with its associated eigenvalue. The  complexity of the figure is only apparent. The plot appears divided into sectors because of the particular numbering employed for the sites. While for a 1D lattice each site can be numbered in an unambiguous manner, and one can hang the plot of the eigenmodes one behind the other, in our case, the geometry is not quite 1D, and we have labelled all sites of the first row as $1,2,\cdots P$. The sites on the second (diluted) row were labelled $P+1, P+2, \cdots N$. An example of this numbering is shown at the top of Fig.\ref{fig:modos_stub_chokley.eps}. Thus, on our plot the amplitudes for the first $85$ sites or so, correspond to the amplitudes on the first row, while the amplitudes for sites $85- 130$ correspond to the amplitudes on the second row of the stub ribbon. 

The modes between $n=45$ and $n=85$ belong to the degenerate eigenvalue $\Omega^2=1$ and they are highly localized and form a Stark-like ladder\cite{stark} with each mode being shifted by one lattice site. The rest of the modes belonging to the dispersive bands, show extended states as usual. In addition we observe the existence of edge modes (marked by red circles), whose shape is shown in the lower part of the figure, where each color denotes the value of the charge residing at a particular ring. Thus, green corresponds to $q_{n}=0$, yellow to $q_{n}=1$ and black to $q_{n}=-1$.
%%%%%%%%%%%%%%%%%%%%%%%%%%%%%%%%%%%%%%%%%%%%
%%%%%%%%%%%%%%%%%%%%%%%%%%%%%%%%%%%%%%%%%%%

\begin{figure}[t]
\begin{center}
\includegraphics[width=8cm]{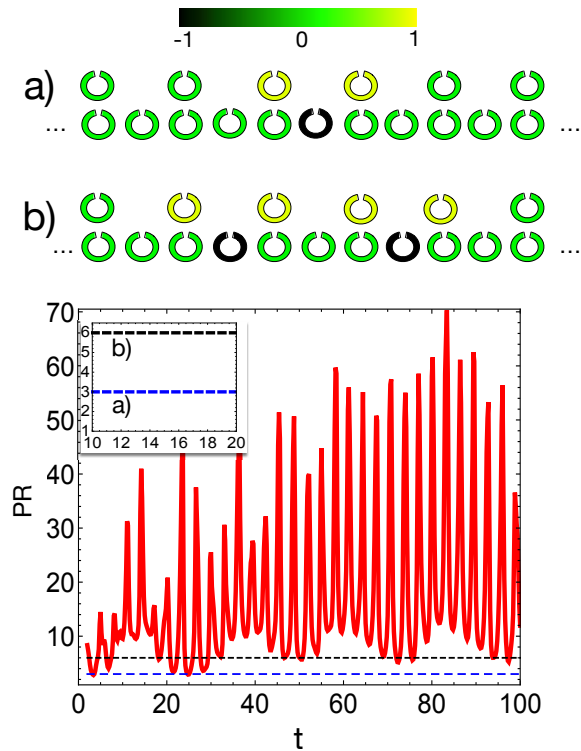}
\end{center}
\caption{ Top: Example of the two lowest-lying compact modes of the stub lattice. Bottom: Time evolution of the participation ratios (PR) of the two compact modes and comparison with the case of an initially localized profile. The inset shows a magnification of the PR of the two modes.
\label{fig:ModoCompacStubRP}}
\end{figure}

The localized modes belonging to the flat band can be combined to yield compacton-like eigenmodes. Two examples of this are shown at the top of figure \ref{fig:ModoCompacStubRP}a,b. The evolution of these two modes is shown in the bottom half of the figure, which shows the participation ratio (PR) vs time. The PR is a measure of localization of a state and is defined as
 $PR=
(\sum_i |q_{i}(t)|^2)^2/\sum_i|q_i(t)|^4$. Roughly, it measures the number of sites in a lattice that are effectively excited by a mode. For a completely localized mode on a single site, $PR=1$, while for a completely delocalized state, $PR=N$, where $N$ is the number of lattice sites. In our case we see that the PR of the two modes does not change with time. As a contrast, we also plot the PR of an initial profile consisting of a single site initially charged. In this case, as time increases the excitation expands in space  filling eventually the whole SRR array.
%%%%%%%%%%%%%%%%%%%%%%%%%%%%%%%%
%%%%%%%%%%%%%%%%%%%%%%%%%%%%%%%

\subsection{dynamical stability}

\begin{figure}[H]
\begin{center}
\includegraphics[width=7cm]{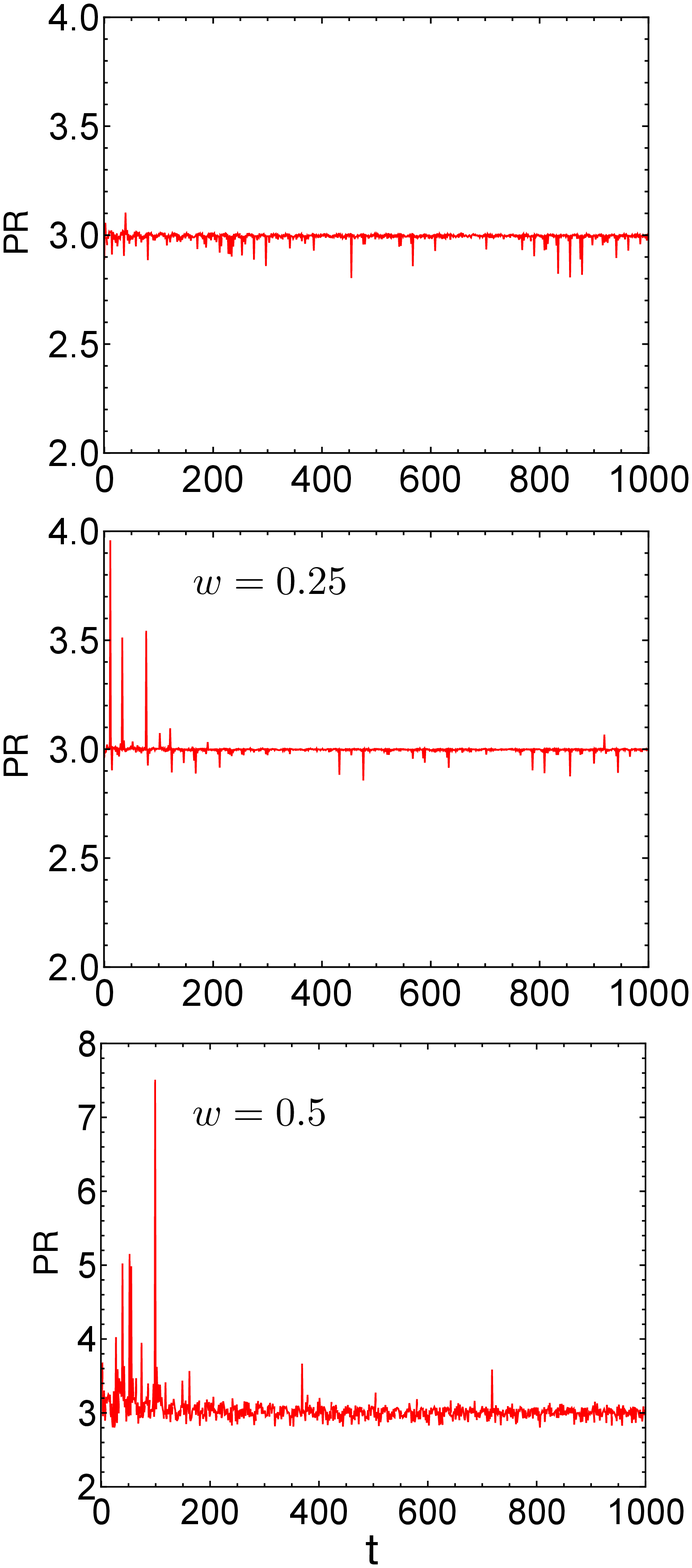}
\end{center}
\caption{ Participation ratio $PR$ of the fundamental mode of the stub ribbon as a function of $t$ for a random perturbation of the initial conditions. The width of the noise is $w=0.1$(top), $w=0.25$ (middle) and $w=0.5$ (bottom).
\label{fig:RandomNoiseStub}}
\end{figure}

Let us now focus on the three-sites fundamental stub mode (Fig.\ref{fig:ModoCompacStubRP}a) and how its evolution is affected by the presence of several perturbations. First, let us disturb the value of the initial charge on the SRRs: $q_j = A_j+\delta_j$, where $A_j = \pm1$ and  $\delta_j\in[-w,w]$, where $2w$ is the disorder width. Figure \ref{fig:RandomNoiseStub} shows the evolution of the PR of this perturbed compact mode, for several disorder widths. As we can see, the mode is quite robust, even for large disorder widths and only for $t<100$ there are strong fluctuations signaling the redistribution of charge among the SRRs. Eventually, the system converges to the non-disordered compact mode, plus some small oscillations. The general features of the evolution in this case can be traced back to the fact that the initial condition can be expanded in modes belonging to all the bands, flat and dispersive and thus, the only portion of the disturbed ring that will evolve away from the ring is the noise part, leaving behind the original ring. In this sense, the disturbed ring is stable. This argument applies to any array, so we can predict stability' for all three geometries. 
%%%%%%%%%%%%%%%%%%%%%%%%%%%%%%%%%%%%%
%%%%%%%%%%%%%%%%%%%%%%%%%%%%%%%%%%%%%

\begin{figure}[H]   
\begin{center}
\includegraphics[width=9cm]{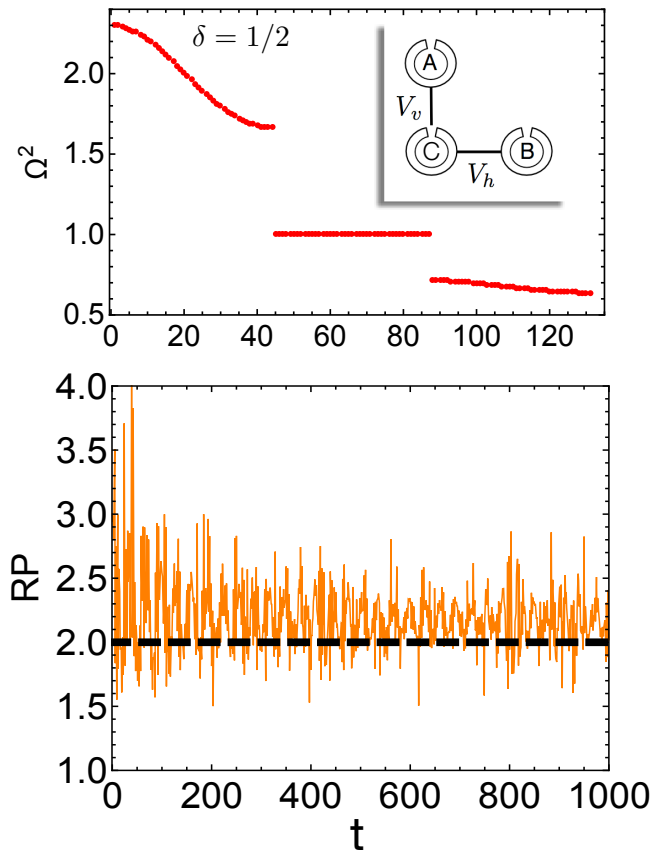}
\end{center}
\caption{Top: Eigenvalues $\Omega^2$ for a long ($N=120$) Stub ribbon for an isotropy parameter $\delta=1/2$. The eigenvalues are plotted in order of increasing value and we show near the flat band region. Bottom: PR of the compact stub mode for an anisotropy parameter $\delta=0.5$.\label{fig:AnisoBandStub} }
\end{figure}

Next, we analyze the effect of a degree of anisotropy. 
We now allow two different couplings: $\lambda_{h}$, the coupling between nearest neighbors along the horizontal direction, and $\lambda_{v}$, the coupling between nearest neighbors in the vertical direction. In a real system this could be due to a uniform stretching of the array, if it is lying on a substrate.  Let us define an anisotropy parameter as $ \displaystyle \delta= \lambda_h/\lambda_v$. From Fig.\ref{fig:AnisoBandStub} we see that a degree of anisotropy does not destroy the flat band.
%%%%%%%%%%%%%%%%%%%%%%%%%%%%%%%%%%%%%%%%%%%%%%
%%%%%%%%%%%%%%%%%%%%%%%%%%%%%%%%%%%%%%%%%%%%%%%%%
The evolution of the PR of the perturbed compact mode, shows that it just oscillates around a constant value,  but it does not decay.  In this case, the oscillation is about $PR \sim 2$, instead of $3$, which is due to the fact that for $\delta = 0.5$ the horizontal sites are farther apart and so, more charge needs to accumulate at the middle site in order to maintain the phase relation necessary for a compact mode to persist. 
%%%%%%%%%%%%%%%%%%%%%%%%%%%%%%%%%
\begin{figure}[h]
\begin{center}
\includegraphics[width=8cm]{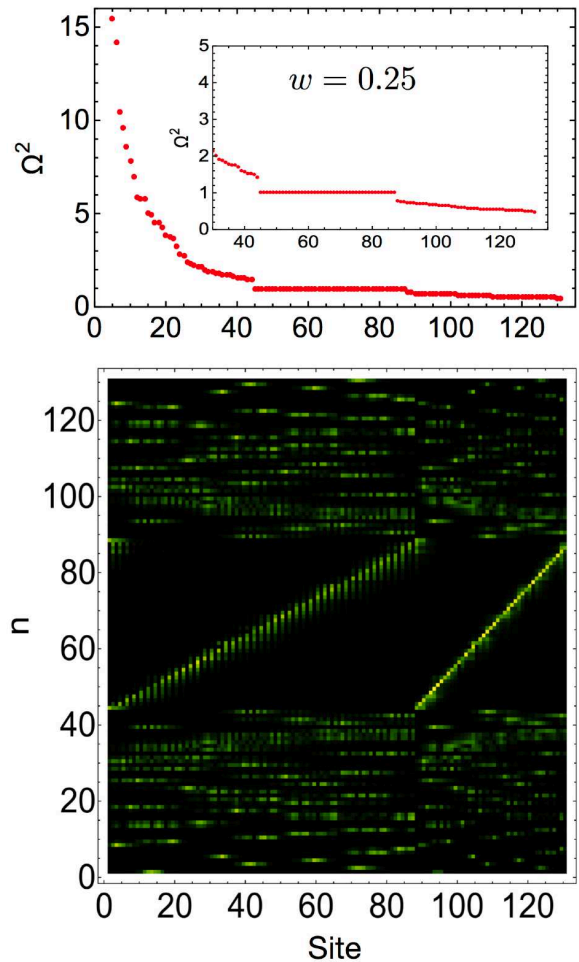}
\end{center}
\caption{Stub ribbon with coupling disorder width $w=0.25$ and $\lambda=0.3$. Top: Eigenvalues of the Stub ribbon, ordered according to increasing value. Bottom: Density plot of the eigenvectors of the Stub ribon, arranged in order of increasing eigenvalue.
\label{fig:RuidoAcopl_Sutb} }
\end{figure} 
%%%%%%%%%%%%%%%%%%%%%%%%%%%%%%%%%%%
Given that the PR is stable, we can estimate the value around which the PR oscillates as follows: By imposing the condition that under the new couplings, the charge is redistributed in such a way as to satisfy the interference condition, one obtains 
\be 
PR(\delta) = {(2\delta^2+1)^2\over{2\delta^4+1.}}
\ee
For $\delta=1/2$ this gives $2$ which almost coincide  with the time average of the PR. 

\begin{figure}[t]   %NO DEBE OMITIR LA H (con mayúscula), de lo contrario la figura no aparecerá.
\begin{center}
\includegraphics[width=9cm]{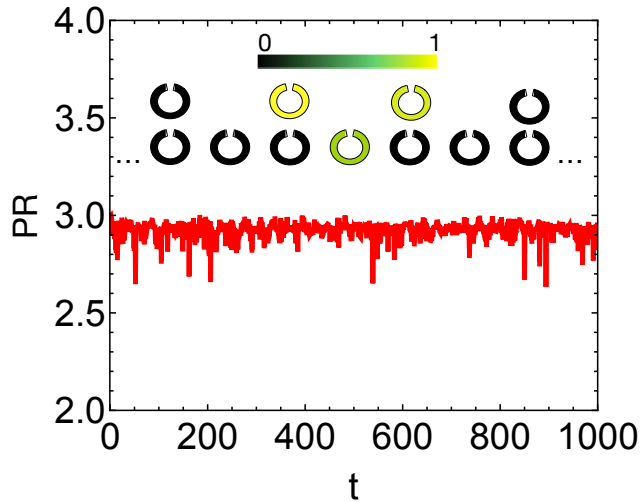}
\end{center}
\caption{Evolution of the participation ratio PR for the fundamental compact mode of the stub ribbon  subjected to coupling noise ($\lambda=0.3, w=0.25$). The inset shows the shape of the compact mode after a long evolution time ($t=1000$).
\label{fig:Ruido_acoplo_RP_stud} }
\end{figure}

Now, let us consider a different and more interesting  type of perturbation: Noise in the mutual couplings. This is given by $\lambda_{i j}\rightarrow \lambda_{i j} + \delta \lambda_{i j}$ with $\lambda_{i j}=\lambda$ and $\delta \lambda_{i j} \in [-w,w]$.
In practice, this type of perturbation could be due to a random variation of the array inter distances coming from a defectively built array. As we can see from Fig.\ref{fig:RuidoAcopl_Sutb}, the inclusion of coupling noise does not destroy the flat band, but   strongly affects the modes in the dispersive bands. These modes show now the phenomenon of localization. This is nothing else than off-diagonal Anderson localization. It is quite visible here since we have chosen a large value for the disorder width, $w=0.25$. The degenerate modes with $\Omega^2=1$ still show the stark ladder-like spatial distribution.
The dynamical evolution of the PR shown in Fig.\ref{fig:Ruido_acoplo_RP_stud} shows that the PR is stable, albeit with oscillations. The fact that the PR never surpasses de value of $3$ when oscillating tell us that the charge is being distributed back and forth among the three SRRs only. Given the large amount of disorder involved ($w=0.25$), the robustness of the dynamical evolution is impressive.
%%%%%%%%%%%%%%%%%%%%%%%%%%%%%%%%%%%%%%%%%%
%%%%%%%%%%%%%%%%%%%%%%%%%%%%%%%%%%%%%% 

\section{The Lieb and kagome ribbons}

For the Lieb and kagome geometrical arrays we repeat exactly the same procedure as the one carried out for the stub lattice (section \ref{stub}). These arrays are shown in Fig\ref{fig:Arreglos} b) and c). We shall proceed summarizing the main results for these two arrays.

\begin{figure}[H]
\begin{center}
\includegraphics[width=8.5cm]{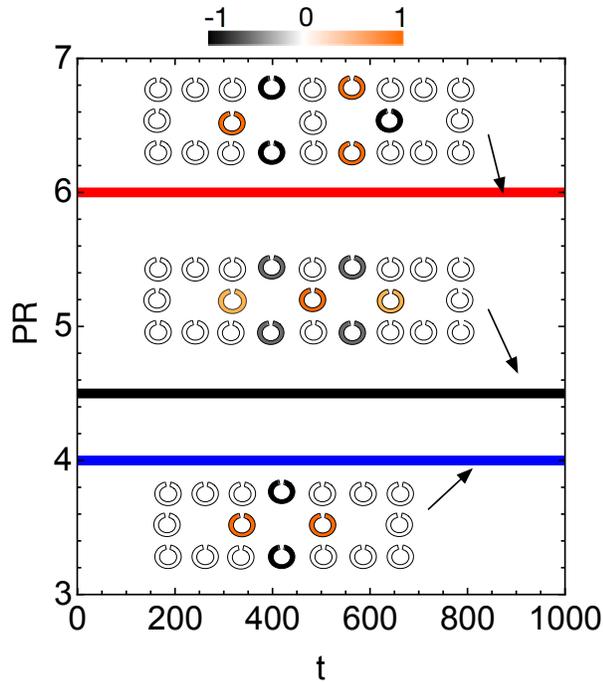}
\end{center}
\caption{ Participation ratio as a function of time
for several combinations of Lieb compact modes. The PR=4 value corresponds to the fundamental mode. The case PR=6 (4.5) corresponds to a linear combination of two fundamental modes in phase (antiphase).
\label{fig:AutovecLieb} }
\end{figure}

\begin{figure}[H]
\begin{center}
\includegraphics[width=8.5cm]{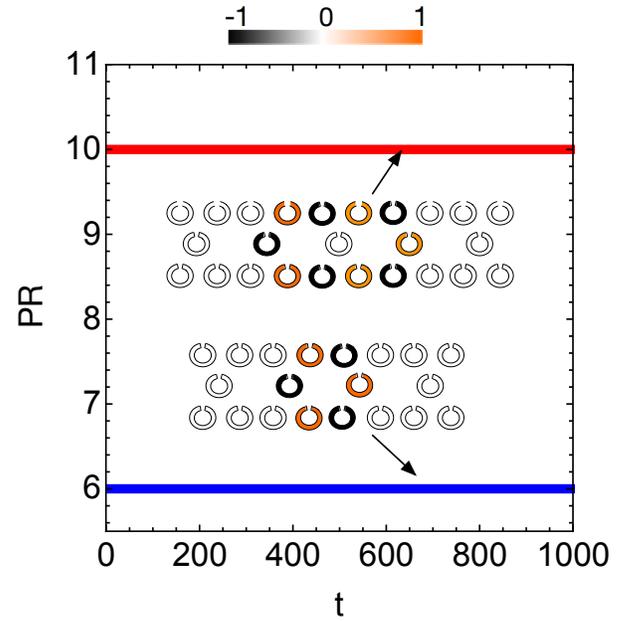}
\end{center}
\caption{ 
Participation ratio as a function of time
for two combinations of kagome compact modes. The PR=6 value corresponds to the fundamental mode, while the PR=10 corresponds to a linear combination of two fundamental modes in antiphase.
\label{fig:AutovecKagome} }
\end{figure}

The Lieb ribbon can be considered as a depleted square lattice ribbon. Its unitary cell contains five sites (SRRs), implying an spectrum consisting of five bands:
\begin{eqnarray}
\Omega^2 &=& 1\nonumber\\
\Omega^2 &=& \frac{1}{1\pm \sqrt{2 (1+\cos(2 k))}\lambda }\nonumber\\
\Omega^2 &=& \frac{1}{1 \pm \sqrt{4 + 2 \cos(2 k)}\lambda} .
\end{eqnarray}
We have a flat band at $\Omega^2=1$. The bands are real provided the coupling satisfies $|\lambda|<1/\sqrt{6}=0.408$. On the other hand, the kagome ribbon also has five units (SRRs) in its unitary cell and its five bands are given by
\begin{eqnarray}
\Omega^2 &=& \frac{1}{1-2\lambda}\nonumber\\
\Omega^2 &=& \frac{1}{1\pm \sqrt{2 (1+\cos(2 k))}\lambda }\nonumber\\
\Omega^2 &=& \frac{1}{1 +\left(1\pm \sqrt{3 + 2 \cos(2 k)}\right)\lambda},
\end{eqnarray}
and shows a $\lambda$-dependent flat band $\Omega^2=1/(1-2\lambda)$. Again, to ensure the existence of the five real bands, $-1/(1+\sqrt{5})<\lambda<1/2$.
%%%%%%%%%%%%%%%%%%%%%%%%%%%%%%%%%%%%%%
\begin{figure}[H]
\begin{center}
\includegraphics[width=7.5cm]{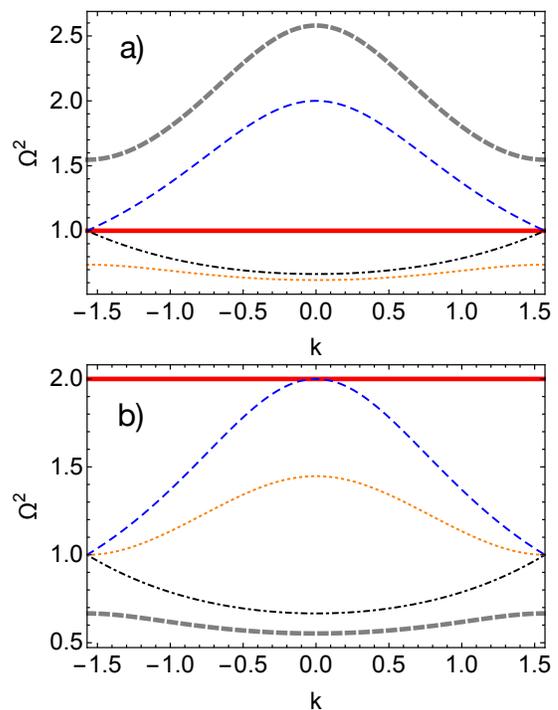}
\end{center}
\caption{ Top: The five bands of the Lieb ribbon. Bottom: The five bands of the Kagome ribbon. The coupling value is $\lambda=0.5$ in both cases.   \label{fig:BandaLiebKagome} }
\end{figure}
%%%%%%%%%%%%%%%%%%%%%%%%%%%
The band structure of the Lieb and kagome ribbons are shown in Fig.\ref{fig:BandaLiebKagome}. 
The eigenmodes belonging to the flatbands of the Lieb and kagome ribbons are also highly localized, in a manner quite similar to the stub case (Fig.\ref{fig:modos_stub_chokley.eps}) and ref\cite{Dany}. Combinations of these modes give rise to compact modes in the form of closed rings, as the ones shown in the inset of Fig.\ref{fig:AutovecLieb} for the Lieb case, and in Fig.\ref{fig:AutovecKagome} for the kagome case. The fundamental Lieb mode is composed of four SRRs forming a ring. Two of them have charge $q$ and the other two charge -$q$. In the kagome case, the fundamental mode is composed of six SRRs, where three of them have charge $q$ while the rest have charge -$q$. The value and sign of the charges are chosen as to ensure that the sites outside the rings will receive no charge at any time, thus isolating the fundamental mode and ensuring its lack of dynamical evolution.
%%%%%%%%%%%%%%%%%%%%%%%%%%%%%%%%%%%%%%%%%%%%
%%%%%%%%%%%%%%%%%%%%%%%%%%%%%%%%%%%%%%%%%%
%%%%%%%%%%%%%%%%%%%%%%
In-phase and out-of-phase combinations of compact modes are degenerate stationary modes and therefore, they also do not diffract. 

\begin{figure}[H] 
    \begin{center}
    \includegraphics[width=7cm]{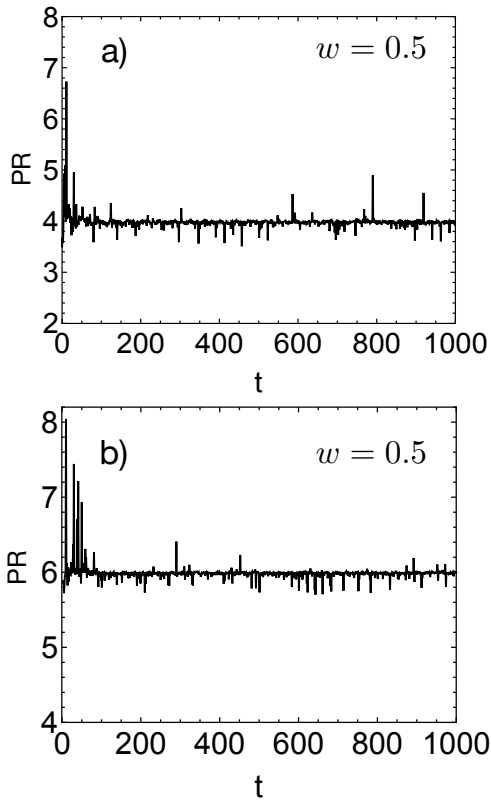}
    \end{center}
    \caption{ Evolution of the participation ratio of the fundamental a) Lieb and b) kagome compact modes
with an initial noise. ($w=0.5$).  \label{fig:RP_kagome_lieb_ruido} }
\end{figure}

As we did for the stub ribbon, let us now look at the dynamical evolution of the fundamental compact modes under the influence of perturbations such as noise in the initial charge, anisotropy of the couplings and noise in the individual couplings. Figure \ref{fig:RP_kagome_lieb_ruido} shows that both ribbons are robust against noise in the initial value of the charges.
%%%%%%%%%%%%%%%%%%%%%%%%%%%%%%%%%
%%%%%%%%%%%%%%%%%%%%%%%%%%%%%%%%%%%%%%%%%%%% 

\begin{figure}[H]   
\begin{center}
\includegraphics[width=9cm]{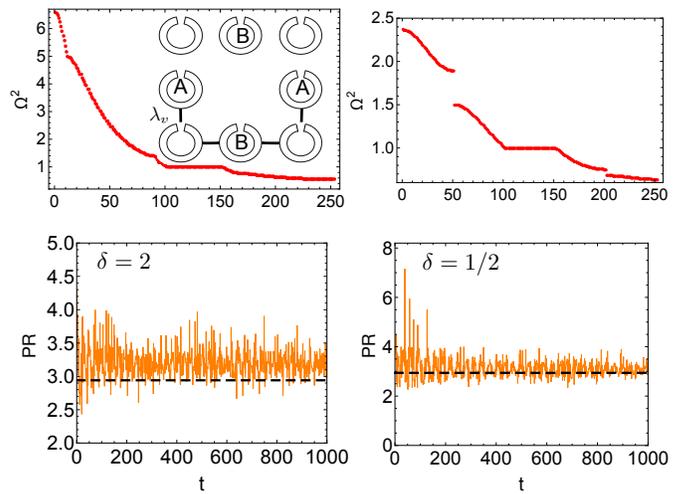}
\end{center}
\caption{ Top: Eigenvalues of the Lieb ribbon, ordered ac- cording to increasing value, for two values of the anisotropy parameter. Bottom: Time evolution of the participation ratio of the Lieb compact mode for two anisotropy parameter values. The horizontal dashed lines mark the theoretical estimate for the PR of the deformed compact modes.\label{fig:RP_Lieb_anisotro_acop}}
\end{figure}

Next, let us examine the stability against a degree of anisotropy in the couplings. Now we have to define two  anisotropy parameters. For the Lieb lattice, $\delta_{l}=\lambda_{h}/\lambda_{v}$, where $\lambda_{h}$ is the coupling between nearest neighbors in the horizontal direction and $\lambda_{v}$ is the coupling between nearest neighbors in the vertical direction. For the kagome lattice we define $\delta_{k}=\lambda_{h}/\lambda_{d}$, where $\lambda_{h}$ is the coupling between nearest neighbors in the horizontal direction while $\lambda_{d}$ is the coupling along the diagonal direction. As we can see from Fig.\ref{fig:RP_Lieb_anisotro_acop}, the flat band of the Lieb ribbon is robust against coupling anisotropy for several $\delta$ values, including large values compatible with the existence of the flat band. The evolution of the participation ratio of the compact Lieb mode ($4$ sites) show that it is stable, with  oscillations around a new equilibrium value. As we did for the stub lattice, we estimate the equilibrium PR by imposing the condition that under the new couplings, the charge is redistributed in such a way as to satisfy the interference condition. Thus,  one obtains the estimate
\be
PR(\delta)=2{(1+\delta^2)^2\over{1 + \delta^4}}.\label{delta}
\ee
For the cases shown in Fig.\ref{fig:RP_Lieb_anisotro_acop}, this estimate predicts that for $\delta=2$ the new stable PR should be $PR=2.94$, while for $\delta=1/2$, the PR should also be $2.94$, exactly as shown in the figure. This is not a coincidence, since $PR(\delta)$ satisfies the symmetry $PR(\delta) = PR(1/\delta)$.
%%%%%%%%%%%%%%%%%%%%%%%%%%%%%%%%%%%%%%%%%%%%%
%%%%%%%%%%%%%%%%%%%%%%%%%%%%%%%%%%%%%

\begin{figure}[H]   
\begin{center}
\includegraphics[width=9cm]{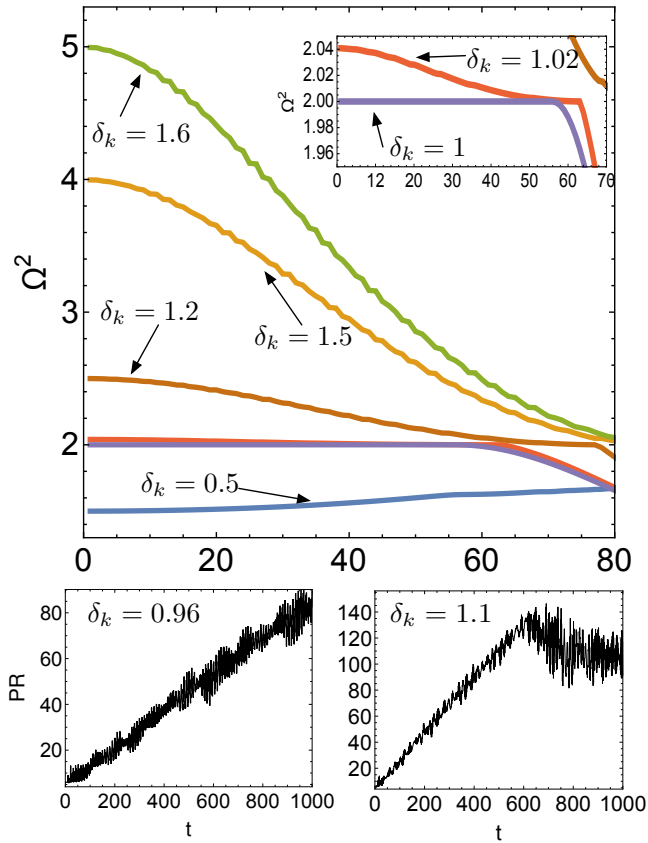}
\end{center}
\caption{Top: Eigenvalues of the kagome ribbon for different anisotropy values. The inset shows the region where one has a flat band when $\delta=1$. Bottom: Evolution of the participation ratio for the compact mode of the kagome ribbon for two anisotropy values.
\label{fig:RP_Kagome_anisotro_acop}}
\end{figure}

A radically different situation occurs for the kagome ribbon\cite{vicencio}. As we can see from Fig. \ref{fig:RP_Kagome_anisotro_acop}, as soon as the anisotropy value is different from one, the flat band is destroyed, even for very small anisotropy values.  When one propagates the compact kagome mode, it deforms quickly spreading its charge all over the lattice, as shown by the evolution of its PR. The oscillations that appear for $t>600$ are due to finite-size effects, where the magnetoinductive wave bounces from the array boundary.
%%%%%%%%%%%%%%%%%%%%%%%%%%%%%%%%%%%%%%%%%
%%%%%%%%%%%%%%%%%%%%%%%%%%%%%%%%%%%%%%
Finally, let us consider a random perturbation of the individual couplings, as we did for the stub lattice: $\lambda_{i j}\rightarrow \lambda_{i j} + \delta \lambda_{i j}$, where $\lambda_{i j}=\lambda$ and $\delta \lambda_{i j}\in [-w,w]$. 
%%%%%%%%%%%%%%%%%%%%%%%%%%%%%%%%%%%%%%
\begin{figure}[t]  
\begin{center}
\includegraphics[width=9cm]{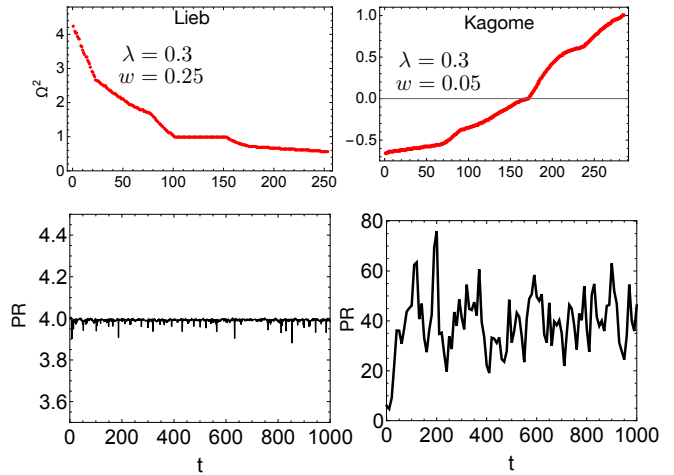}
\end{center}
\caption{Top: Eigenvalue spectrum for the Lieb ribbon (left) and the kagome ribbon (right), for $w=0.25$ (Lieb) and $w=0.05$ (kagome). Bottom: Evolution of the participation ratio of the Lieb (left) and kagome (right) fundamental compact modes.
\label{fig:Lieb_kagome_ruido_v2} }
\end{figure}
%%%%%%%%%%%%%%%%%%%%%%%%%%%%%%%%%%%%%%%%%%
Results are shown in Fig.\ref{fig:Lieb_kagome_ruido_v2}. For the Lieb ribbon we can see that the flat band survives the addition of coupling noise, even for large disorder strength ($w=0.25$) while for the kagome ribbon, the flat band is destroyed even for very small disorder strength ($w=0.05$). The evolution of the PR for the Lieb ribbon shows that it is quite robust with a PR value around the one corresponding to the fundamental mode ($PR=4$), while for the kagome case, the PR quickly reaches ``macroscopic'' values, of the order of $N$.
%%%%%%%%%%%%%%%%%%%%%%%%%%%%%%%%%%%%%%%%%%%%%
\setlength{\tabcolsep}{3em}
\begin{table*}[t]
\begin{tabular}{  l c l c l c}
\hline
    \hline
    Ribbon & Initial noise & Anisotropy & Coupling Noise \\ \hline 
    Stub & Stable & Stable & Stable\\ \hline 
    Lieb & Stable  & Stable & Stable\\ \hline 
    Kagome & Stable & Unstable & Unstable  \\ \hline 
    \hline
\end{tabular}
\caption{Summary of the stability behavior of the three ribbons against the three types of perturbation.\label{tabla1}}
\end{table*}
%%%%%%%%%%%%%%%%%%%%%%%%%%%%%%%%%%%%%%%%%%%%%%%
To summarize the stability results for all three ribbons, we show in Table 1 the response of each lattice to the three types of perturbations considered in this work. While the Stub and the Lieb ribbons are stable to all the perturbations, the kagome was only stable to the initial noise only, being unstable against the other two. Clearly, there is a strong geometrical element at play here.

\section{Dissipation}

It is well-known that in SRRs some degree of loss (radiative and resistive) is always present. It is natural then, to wonder whether our flatbands survive the addition of dissipation. In the presence of lossses, the equation for the stationary modes changes to
\be
-\Omega^2\left(q_n+\lambda \sum_{nn} q_m\right) + i \gamma \Omega q_n + q_n=0.
\label{eq:NewEstacionaria}
\ee
where $\gamma$ is the loss coefficient. By means of elementary algebraic manipulations, one can recast Eq.(\ref{eq:NewEstacionaria}) as
\be
\beta q_n + \lambda \sum_{m\neq n} q_m = 0
\ee
i.e., the well-known tight-binding equation, where the eigenvalue $\beta$ is given by 
\be
\beta = 1-\left({1 + i \gamma \Omega \over{\Omega^2}}\right)\label{new}
\ee
The mapping defined by Eq.(\ref{new}) allows us to transit between our SRRs ribbons with dissipation, and the tight-binding ribbons, of the same geometry. We immediately see that the flatbands of the tight-binding system imply flatbands on the SRR system, which are complex as soon as the loss coefficient is included.
That is, the flatbands without losses 'survive', meaning they do not become dispersive, but they become complex. The flatbands of the tight-binding ribbons were examined by us in a previous work\cite{previous}. These are $\beta=0$ (stub and Lieb), $\beta= -2 \lambda$ (kagome). From Eq.(\ref{new}), we obtain the flatbands of the SRRs with losses:
\be
\Omega=(1/2)\left( i \gamma + \sqrt{4-\gamma^2} \right)\label{eq:stubandlieb}
\ee
(stub and Lieb),
\be
\Omega= {i\  (\gamma/2)\over{1 - 2 \lambda}} +\sqrt{{1\over{1 - 2 \lambda}}- {(\gamma/2)^2\over{(1-2 \lambda)^2}}}\label{kagome}
\ee
(kagome).
Comparison between Eqs.(\ref{eq:stubandlieb}) and (\ref{kagome}), shows that 
dissipation effects are stronger in the kagome lattice than in the stub and lieb lattices.

\section{Conclusions}

We have examined the spectral and dynamical properties of three quasi-onedimensional (ribbons) magnetic lattices that have a flatband in their spectra. It is observed that the degenerate eigenmodes of these flatbands form a Stark-like ladder where each mode is shifted by one lattice site. Their combination gives rise to compact modes that do not diffract due to a geometrical phase cancellation. For all three cases we computed the stability of the fundamental band mode against perturbation of their initial charge value, the effect of possible anisotropy of the couplings, and the presence of small random perturbations of the individual couplings. We find that the Stub and Lieb ribbons are stable against all three perturbations, while the kagome ribbon is only stable to the perturbation of the initial value.  At this point,  it is still an open question whether there is a ribbon of adequate geometry whose compact fundamental mode be stable against all three types of perturbation considered in this work. The introduction of dissipative effects maintains the non dispersive character of the flatbands, but they become complex for any dissipation strength with the kagome ribbon being the most susceptible to losses. We can conclude that the geometry of the magnetic array plays an important role in determining the stability and loss rate of the flatband modes.

\section{Acknowledgements}

This work was partially supported by FONDECYT grants 1120123 and 1160177 and  Programa ICM P10-030-F.

\end{document}